\documentclass[aps,prl,reprint,superscriptaddress]{revtex4-1}

\usepackage{graphicx}
\usepackage{dcolumn}
\usepackage{bm}
\usepackage{color}

\begin{document}

\title{Direct measurement of cell wall stress--stiffening and turgor pressure in live bacterial cells}

\author{Yi Deng}
\affiliation{Department of Physics, Princeton University, Princeton, NJ 08544, USA}
\author{Mingzhai Sun}
\affiliation{Lewis-Sigler Institute for Integrative Genomics, Princeton University, Princeton, NJ 08544, USA}
\author{Joshua W.~Shaevitz}
\email{shaevitz@princeton.edu}
\affiliation{Department of Physics, Princeton University, Princeton, NJ 08544, USA}
\affiliation{Lewis-Sigler Institute for Integrative Genomics, Princeton University, Princeton, NJ 08544, USA}

\date{\today}

\begin{abstract}
We study intact and bulging \textit{Escherichia coli} cells using atomic force microscopy to separate the contributions of the cell wall and turgor pressure to the overall cell stiffness. We find strong evidence of power--law stress--stiffening in the \textit{E. coli} cell wall, with an exponent of $1.22\pm0.12$, such that the wall is significantly stiffer in intact cells ($E = 23\pm8$~MPa and $49\pm20$~MPa in the axial and circumferential directions) than in unpressurized sacculi. These measurements also indicate that the turgor pressure in living cells \textit{E. coli} is $29\pm3$~kPa.
\end{abstract}

\maketitle

Many cellular-scale processes in biology, such as cell growth, division and motility, necessarily involve mechanical interactions. Recent theoretical work in bacteria has led to a number of physically--realistic models of bacterial cells \cite{Furchtgott2011,Sun2010, Jiang2010}. However, in many instances, precise, direct measurements of the mechanical properties of cellular components in live cells are lacking.

The cell envelope in most bacteria is made of one or two layers of membrane and a rigid cell wall consisting of a network of peptidoglycan (PG) polymers. These two materials serve different cellular functions. The semi-permeable plasma membrane maintains a chemical separation between the cell interior and the surrounding medium. The large concentration of solutes in the cytoplasm generates an osmotic pressure, termed turgor pressure, that pushes the plasma membrane against the cell wall. The cell wall, on the other hand, defines the cell shape and constrains the volume under turgor.

The magnitude of the turgor pressure under physiological conditions has been estimated using several techniques: by collapsing gas vesicles in rare species of bacteria \cite{Holland2009}, by AFM indentation \cite{Arnoldi2000, Yao2002}, and by calculating the total chemical content of the cytoplasm \cite{Cayley2000}.  The estimated pressure values vary by more than an order of magnitude, from $10^4$ to $3\times 10^{5}$~Pa. While mechanical experiments, such as AFM indentation, are the most direct probes, separating the mechanical contributions of the wall and pressure has not been previously possible and thus these experiments may only provide an upper bound on the true turgor pressure.

Similarly, the elasticity of the cell wall has been difficult to probe in individual, live cells. Most previous mechanical measurements on the cell wall have been performed using chemically isolated walls, termed sacculi, that may be altered from the native state, or on large bundles of cells \cite{Mendelson2000}. Yao \textit{et~al.} reported an anisotropic elasticity of $25$ MPa and $45$ MPa in the axial and circumferential directions relative to a cell's rod-shape using single  flattened \textit{E.~coli} sacculi adhered to a substrate \cite{Yao1999}. Thwaites and coauthors probed the elastic modulus of macroscopic threads of many \textit{Bacillus subtilis} sacculi in humid air and found that the modulus varied from $10$ to $30$ MPa depending on the humidity and salt concentration \cite{Thwaites1985, Thwaites1989, Mendelson1989}.  Attempts to probe whole--cell elasticity have also been made using AFM indentation of \textit{Myxococcus xanthus} cells \cite{Pelling2005} and optical--tweezer bending of  \textit{Borrelia burgdorferi} sacculi \cite{Dombrowski2009}.

In addition, because the PG material is essentially a cross-linked polymer mesh, it is expected to exhibit a substantial amount of stress--stiffening \cite{Gardel2004,Koenderink2006,Tharmann2007, Broedersz2008, Kim2009}. Unpressurized sacculi thus provide a poor platform for estimating the wall elasticity in live cells. Boulbitch \textit{et.~al.} modeled the cell wall as a deformable hexagonal mesh and predicted a load-dependent elasticity with a power--law stress--stiffening exponent of about one \cite{Boulbitch2000}. Thwaites and coauthors found about an order of magnitude change in the thread modulus upon loading, although it is unclear how to interpret measurements from these very large, multi--sacculus objects performed in air \cite{Thwaites1985, Thwaites1989, Mendelson1989}.

\begin{figure}[thb]
\includegraphics[width=0.45\textwidth]{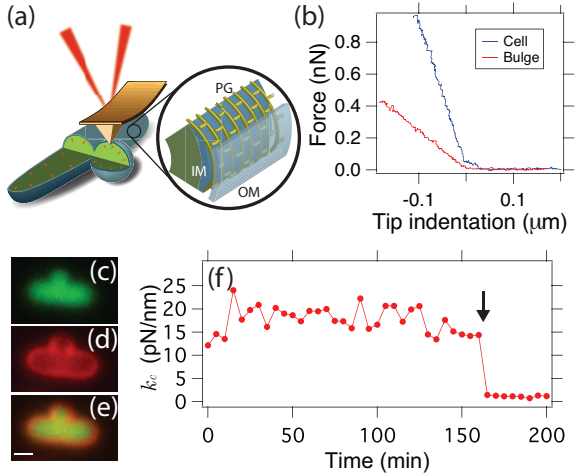}
\caption{\label{fig_intro}(a) Schematic cartoon illustrating the bulging \textit{E.~coli} and AFM stiffness measurement.  The magnified region shows the details of the inner membrane (IM), peptidoglycan (PG) network and the outermembrane (OM).  (b) Typical force-indentation traces obtained by indenting a cell and bulge.  (c) Cytoplasmic GFP (green) is able to occupy both the cell and bulge interiors, indicating the ability of protein--sized objects to transverse the pore. (d) A membrane stain (FM4-64, red) labels the outer membrane.  (e) Overlay of the cytoplasmic EGFP and membrane stain.  Scale bar is 1 $\mu m$.  (f) Cell stiffness shows little variation before the bulging event (arrow) at which point it drops suddenly.}
\end{figure}

Mechanical indentation of live cells is likely the most direct method for probing these sorts of mechanical properties. Under external perturbation, however, the cell wall and turgor pressure have mixed contributions to the response, making it hard to independently estimate these two quantities.  By studying a bulging strain of  \textit{E.~coli}, we are able to simultaneously determine both the wall elasticity and the turgor pressure and reveal their dependence [Fig. \ref{fig_intro}].

Briefly, we first obtain the turgor pressure of individual bulging cells from the  bulge radius and indentation stiffness using AFM and fluorescence microscopy [Fig. \ref{fig_intro}(a-e)]. Then, from the size and stiffness of the cell body, we are able to extract the elasticity of the cell wall under tension using numerical methods. The variation in turgor pressure among bulging cells allows us to probe the mechanical properties of the PG over a broad range of stresses.   Additional experiments using  non--bulging cells yields the turgor pressure and wall modulus of \textit{E.~coli} under physiological conditions. More details regarding the experimental procedures are described in the supplemental materials \footnote{See supplemental material at http://link.aps.org/supplemental/... for details regarding experimental methods and theoretical derivations.}.

Several lines of evidence indicate that the cell wall in bulged cells is not significantly different than in non--bulged cells. First, bulging is a discrete event that is completed within a few seconds. Second, the cell stiffness remains constant in the presence of vancomycin until the sudden bulging event when the stiffness drops dramatically [Fig. \ref{fig_intro}(f)].  Taken together, these indicate that the mechanical properties of the cell wall as a whole are unaffected by drug treatment except at the precise location of fracture and bulging.

GFP molecules are able to move between the bulge and the cell interiors [Fig. \ref{fig_intro}(c)], indicating that cytoplasmic objects smaller than at least 3--4~nm are free to exchange between these compartments. Because the turgor pressure overwhelmingly results from the concentration of small solutes, the pressure in the cell and bulge can be considered the same. We calculate this pressure from the stiffness of the bulge by modeling the bulge as a liquid vesicle and the shape of the AFM tip as a cone \footnotemark[1].

\begin{figure}
\includegraphics[width=0.45\textwidth]{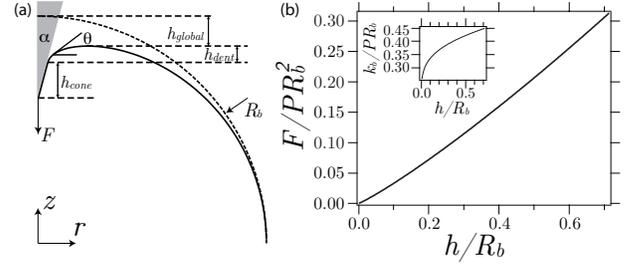}
\caption{\label{fig_bubble}Model of a fluidic membrane bulge under a force $F$ exerted by a conical indenter.  (a) The total deformation of the bulge consists of a global deformation, $h_{global}$, a local dent $h_{dent}$ and the contact height $h_{cone}$.  The dashed line is a sphere of radius equal to the bulge waist. (b) The dimensionless force--indentation relation is nearly linear.  Inset: dimensionless stiffness vs. indentation.}
\end{figure}

Briefly, the total indentation size for an indentation force $F$,  a bulge of radius $R_b$, an indenter half--conical angle of $\alpha$ and pressure $P$ is given by \footnotemark[1]
\begin{equation}
\begin{array}{l}
h =h_{gobal}+h_{dent}+h_{cone};\\[0.5ex]
h_{global} = R_b-\frac{\sigma_b}{P}\left[1+I(\pi/2, a)\right];\\[0.5ex]
h_{dent} = \frac{\sigma_b}{P}\left[1-\sin\alpha-I(\pi/2-\alpha, a)\right];\\[0.5ex]
h_{cone} = \frac{\sigma_b}{P}\left[(\sqrt{\cos^2\alpha+a}-\cos\alpha)\cot\alpha\right]
\end{array}
\end{equation}
where the bulge surface tension $\sigma_b = PR_b/2 - F/2\pi R_b$, $a= PF/\pi\sigma_b^2$, and $I(\xi, a) = \int_{0}^{\xi} \sin^2\zeta(a+\sin^2\zeta)^{-1/2}d\zeta$. The  indentation, $h$, has a nearly linear dependence on the indentation force [Fig. \ref{fig_bubble}(b)].  Under experimental conditions where $\alpha=\pi/12$, $R_b\sim0.5 \mu$m, $P\sim1$~kPa and $F\sim0.01-0.1$~nN, the dimensionless spring constant $k_b/PR_b$ varies from 0.35 to 0.38 [Fig. \ref{fig_bubble} (b) \textit{inset}].

For each bulging cell, we measure $h / R_{b}$ and use the model to obtain the reduced stiffness $k_b/PR_{b}$ as shown in the inset of Fig. \ref{fig_bubble}(b). From the mechanical measurements of the bulge stiffness and radius, we then calculate the turgor pressure $P$ in that particular cell. We then use this value to estimate the circumferential surface tension experienced by the cell wall, $\sigma_\perp=PR_c$, where $R_c$ is the cell radius \footnote{Circumferential quantities are  denoted as $\perp$ while axial quantities are denoted as $\parallel$.}.

\begin{figure}[htb]
\includegraphics[width=0.5\textwidth]{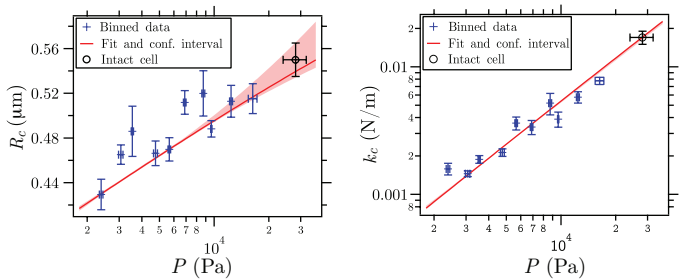}
\caption{\label{fig_Rkvssigma}Bulging cell radius $R_c$ and indentation stiffness $k_c$ are plotted against cell turgor pressure $P$.  Data from 72 bulged cells are binned in 10 logarithmically--spaced bins using weights from the relative error estimates of the individual indentation traces and fluorescent images (blue crosses). Data from 42 non-bulged cells are plotted as black open squares.  Red lines indicate the best fit of the stress--stiffening model along with 68\% confidence intervals.}
\end{figure}

Figure \ref{fig_Rkvssigma} shows the cell radius, $R_c$, and stiffness, $k_c$,  as functions of the pressure derived from bulge indentation.  Both radius and stiffness are positively correlated with the turgor pressure.  We further determined the size and stiffness of non--bulging cells to be $0.55\pm0.02$~$\mu$m and $0.017\pm0.002$~N/m, respectively \footnote{The size and stiffness of a similar strain of \textit{E. coli} that does not carry the \textit{imp4213} mutation was within 10\% of the values for the \textit{imp-} strain, indicating that increased outer membrane permeability does not have a large effect on the turgor pressure or cell wall elasticity.}.

\begin{figure}[htb]
\includegraphics[width=0.4\textwidth]{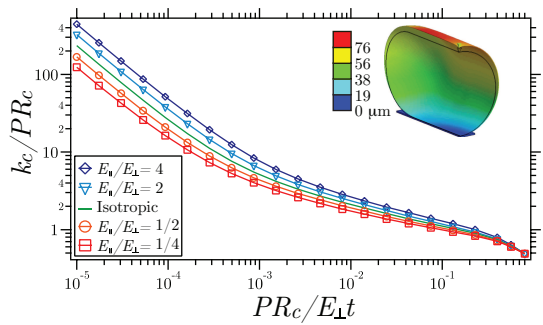}
\caption{\label{fig_kPRvsPREt_ortho} Simulated value of reduced cell indentation stiffness $k_c/PR_c$ against the reduced inflation magnitude $PR_c/E_\perp t$ for different orthotropic ratios of the stretching elasticity in the axial direction to the circumferential directions, $E_\parallel/E_\perp$.  (inset) The result of a single simulation. One quadrant of the indented cylinder is shown, with color labeling the displacement in the indentation direction.  The black wireframe shows the undeformed, unpressurized capsule.}
\end{figure}

The indentation stiffness of the cell wall is governed by terms associated with stretching and bending of the PG as well as terms related to the surface tension. While the bending energy of the wall has been shown to be negligibly small \cite{Arnoldi2000}, we cannot ignore the stretching energy of the PG network and thus analysis of the cell indentation data is more complicated than for bulge indentation. To address this problem, we used finite--element calculations of the force--indentation relation for an inflated cylindrical shell [Fig. \ref{fig_kPRvsPREt_ortho} \textit{inset}].

Rather than attempting to estimate the elastic parameters for each measured cell, we generated a numerical model for the radius, $R_c$, and stiffness, $k_c$, in the presence of stress--stiffening and performed a global fit to all the cellular indentation data. This procedure and the fitting results are sketched below, while in depth derivations and modeling details are provided in the supplemental materials \footnotemark[1].

We incorporate stress--stiffening in the cell wall by describing the nonlinear elasticity of the PG network as a power law in the turgor pressure, $E_\perp = E_0 (P/P_0)^\gamma$. $E_0$ is the Young's Modulus at reference pressure $P_0$ (fixed at 5~kPa, in the middle of the  range of measured bulge pressures), and $\gamma$ is the stress--stiffening exponent. Here, the nonlinearity is only dependent on the pressure in a given cell and we ignore the much smaller change in stress caused by AFM indentation. The independent parameters $\gamma$ and  $E_0t$, where $t$ is the thickness of the cell wall, fully define the nonlinear elasticity. These two quantities, combined with the radius of a cell at the reference pressure, $R_0$, make up the fitting parameters for interpreting bulged cells.  Our global fit additionally includes the radius and stiffness data from the non--bulged, intact cells which introduces one additional free parameter: the physiological turgor pressure.

The radial expansion, $R_c(P; E_{0}t,\gamma, R_0)$, can be solved implicitly from the following equation as derived in the supplemental materials \footnotemark[1]
\begin{equation}
\frac{P}{P_0} =
 \frac{R_0}{R_c}\left[\frac{(\gamma-1)E_0t}{\gamma P_0R_0}\left[(\frac{R_0}{R_c})^\gamma-1\right]+1\right]^{\frac{1}{1-\gamma}}.
\end{equation}
The dimensionless quantity $PR_c/E_\perp t$ describes the magnitude of inflation under pressure.

Calculation of the cell stiffness under pressure, $k_c(P; E_{0}t,\gamma, R_0)$, is significantly more complicated. The dimensionless stiffness, $k_c/PR_c$, depends only on $PR_c/E_\perp t$ as can be found from scaling arguments \footnotemark[1], and monotonically decreases as the cylinder is inflated due to the relative magnitudes of surface tension and shell bending [Fig. \ref{fig_kPRvsPREt_ortho} \textit{green line}]. However, stress stiffening adds an extra complication due to an anisotropy inherent in a cylindrical geometry; the surface tension in the circumferential and axial directions of a cylinder are different by a factor of 2. Therefore, the Young's modulus, which is a function of surface tension, is orthotropic. We simulated indentation of pressurized cylinders with several different values for the elastic anisotropy, $E_\parallel / E_\perp$ [Fig. \ref{fig_kPRvsPREt_ortho}]. For a given pressure, the anisotropy can be calculated \footnotemark[1] and the  correct relationship between the dimensionless stiffness and the radial inflation can be interpolated using the curves shown in Fig. \ref{fig_kPRvsPREt_ortho}. Combined with the radial expansion function, this is sufficient to solve for $k_c(P; E_{0}t,\gamma, R_0)$.

The results of a global fit of the functions $R_c(P; E_0t, \gamma, R_0)$ and $k_c(P; E_0t, \gamma, R_0)$ to the experimental data are shown in Fig. \ref{fig_Rkvssigma}. The best fit yields parameter estimates of $E_0t=0.026\pm0.001$~N/m, $\gamma = 1.22\pm0.12$, $R_0=464.2\pm0.9$~nm and a turgor pressure $P=29\pm3$~kPa. At this turgor pressure, using the estimated cell wall thickness $4.5\pm1.5$~nm \cite{Yao2002}, the  cell wall Young's moduli are $E_\perp = 49\pm20$~MPa and $E_\parallel = 23\pm8$~MPa.

Previous work using AFM indentation of bacteria has been used to quantify turgor pressure and cell wall elasticity \cite{Arnoldi2000, Yao2002}. In that work, the relationship between linear indentation and surface tension was established, but the stretching of the cell wall was neglected or at most underestimated.  Our study, which independently measures the turgor pressure and cell stiffness, suggests that cell wall stretching and surface tension contribute similar amounts to the indentation stiffness. This is most evident in the difference in the $k/PR$ ratio for membrane bulges, $\sim 0.36$, and cells, $\sim 0.9$.  This difference arises from the fluidity of lipid membranes; while the bulge can redistribute material to minimize stress, the rigid cell wall can not.  For the cell wall, therefore, the overall stiffness depends on stretching even in a tension-dominated regime.

Mendelson and others introduced a pressure--independent, tube--bending  method  to  quantify  cell wall elasticity \cite{Mendelson2000}.  Wang \textit{et~al.} bent live \textit{E.~coli} cells and found their flexural rigidity to be $2.0\pm0.4\times10^{-20}$ Nm$^{2}$ \cite{Wang2010}.  This result yields an axial cell wall Young's modulus, including uncertainties in the wall thickness, of $E_\parallel = 11 \pm 4$~MPa, in  agreement with our measurements. Using our numerical model, we combined this value of the axial modulus with the stiffness of intact cells measured using AFM indentation and estimate the turgor pressure in intact cells to be $35 \pm 7$~kPa. This bulge--free measurement further validates our estimate of the turgor pressure and cell wall stress--stiffening.

Polymer networks often exhibit a nonlinear stress-strain relation due to intrinsic geometric nonlinearities and a potential nonlinear force-extension relation of the individual polymers at finite temperature \cite{Gardel2004}.  Boulbitch \textit{et.~al.} modeled the PG network as a hexagonal mesh of rigid glycan subunits and elastic peptide cross-links.  They predicted a power--law relationship between the axial elastic modulus and stress with a stiffening exponent  of $\sim 1$ \cite{Boulbitch2000}.  We find a stiffening exponent of $1.22\pm0.12$ in the \textit{E. coli} cell wall in quantitative agreement with the model and similar to observations from gram--positive \textit{Bacillus} sacculus threads \cite{Thwaites1989}.

To summarize, we used AFM and fluorescent microscopy to probe the elastic properties of live \textit{E.~coli} cells using a system that allows us to separately probe pressure and elasticity.  Our results indicate that the turgor pressure in live cells is $\sim30$~kPa, or $\sim 0.3$~atm. This value is  lower than previous chemical estimates of the pressure but similar to other  mechanical measurements. Our data further indicate that the cell wall stress-stiffens. Stress--stiffening affords a unique mechanical advantage to cells by preventing abrupt cell shape changes during changes in external pressure or osmolarity while maintaining a relatively compliant cell elasticity under normal conditions.

\begin{acknowledgments}
This research was supported by the Pew Charitable Trusts and NSF Award PHY-0844466.  We gratefully thank Natacha Ruiz and Tom Silhavy for help in constructing cell strains and Ned Wingreen for helpful advise.
\end{acknowledgments}

\end{document}

\maketitle

\section{Experimental methods}
The bulging \textit{E.~coli} strain we use is derived from the K12 wild-type strain and contains a mutation, \textit{imp4213}, that increases the outer membrane permeability to allow small molecules to enter the periplasmic space \cite{Eggert2001, Huang2008}. We then use vancomycin, a drug that inhibits PG subunits from forming peptide cross-links, to generate a small number of local fractures in the cell wall.  Under turgor pressure, the cytoplasm pushes the inner membrane through the fracture and forms a membrane bulge outside the cell wall [Fig. 1(a), (c-e)].  In addition to the \textit{imp4213} mutation, we knocked out genes that encode external cellular appendages that interfere with the AFM tip ($fliC$ and $fimA$). Cells also carry plasmid pWR20 which encodes for a moderate level of expression of the fluorescent protein EGFP and kanamycin resistance.

Cells are grown in LB medium containing 50~$\mu$g/ml kanamycin at 37$^\circ$C to OD 0.3, followed by the addition of vancomycin (20~$\mu$g/ml) and a 10 minute incubation.  Cells are then immobilized on  poly-l-lysine (PL) coated glass coverslips.  In the presence of the drug, cells stochastically form bulges along the cell cylinder. We probe the stiffness of the cell and bulge with a custom-built AFM/fluorescence microscope [Fig. 1(a)].

Mechanical stiffness is measured by comparing the slope of indentation on the cell, bulge and glass surface \cite{Arnoldi2000}. To exclude the effect of viscosity on the stiffness, we  tested the stiffness at several indentation speeds and found similar results. All measurements used a pyramidal-tipped cantilever (stiffness = 11 pN/nm, calibrated using the thermal deflection spectrum  \cite{Sader1999}) and an indentation speed of 3~$\mu$m/s. The cell radius is obtained from the point of contact between the tip and the cell. The bulge radius is obtained from fluorescence microscopy [Fig. 1(c)].

\section{Finite element simulations of cell indentation}
In our simulation, the cell wall is modeled as a tube with 2~$\mu$m length, and terminated with spherical endcaps (Fig. 4 \textit{inset}).  The simulation was performed on a quadrant of the endcapped cylinder with symmetric boundary condition.  We adopt the convention of natural, or engineering, stress and strain to define the Young's modulus $E$ and set the poisson ratio to zero. The elastic modulus is set to $20$~MPa,  the thickness to $6$~nm (yielding a combined parameter $Et=.12$~N/m), the  cell radius in the absense of pressure to $500$~nm and the cone angle of the indenter to $\pi/12$ with a spherical tip of radius $7.5$~nm.  The turgor pressure is chosen to be the independent variable, and the indentation stiffness is obtained from the force required to create an indentation of $1/20$ of the cell radius.

\section{Bulge stiffness under conical indenter}
The shape of a deformed, pressurized membrane bulge indented by a conical indenter of half cone--angle $\alpha$ with force $F$ can be  solved analytically.  Let $P$ be the pressure in the bulge and $\sigma_b$ the surface tension. Note that $\sigma_b$ is uniform on the entire liquid membrane bulge.  From the force balance condition in the axial direction, the indentation force in cylindrical coordinates $(r, \phi, z)$ is given by
\begin{equation}
\label{eqn_balance}
\pi P r^2 + 2\pi r \sigma_b \sin\theta = F,
\end{equation}
where $\theta$ is the elevation angle of the bulge tangential direction in the axial cross-section \cite{Yao2002}.  The radial coordinate of the bulge contour $r$ and it's derivative are
\begin{equation}
\label{eqn_r}
r = \frac{\sigma_b}{P}\left(\sqrt{\sin^2\theta + a} - \sin\theta\right),
\end{equation}
and
\begin{equation}
\label{eqn_dr}
\frac{dr}{d\theta} = -\dfrac{r\sigma_b \cos\theta}{Pr+\sigma_b\sin\theta},
\end{equation}
where $a = \dfrac{PF}{\pi\sigma_b^2}$.  At $\theta = -\pi/2$, $r$ reaches the bulge radius $R_b$ and the surface tension $\sigma_b$ can be solved from Equation (\ref{eqn_balance}):
\begin{equation}
\sigma_b = \dfrac{PR_b}{2} - \dfrac{F}{2\pi R_b}
\end{equation}

Substituting Equations (\ref{eqn_r}) and (\ref{eqn_dr}) into $\dfrac{dz}{dr}=\tan\theta$, we obtain
\begin{equation}
\label{eqn_dz}
\begin{array}{lcl}
dz & = & \tan\theta dr\\[2ex]
{}& = & \tan\theta\dfrac{dr}{d\theta}d\theta\\[2ex]
{}& = & \dfrac{\sigma_b}{P}\left( \dfrac{\sigma_b\sin^2\theta}{\sqrt{\sin^2\theta+a}}-\sin\theta\right)d\theta.
\end{array}
\end{equation}
Integrating over $z$, the shape of the bulge is solved as a function of the elevation angle $\theta$.  Here, we separate the total indentation into three parts: $h =h_{gobal}+h_{dent}+h_{cone}$ (Fig. 2). These are the distance from the highest point on the deformed bulge to the undeformed bulge pole, $h_{global}$; the height from the indenter contact point to the highest point on the bulge, $h_{dent}$; and the depth of the contact region between the cone and bulge, $h_{cone}$.  We further define the elliptical integral
\begin{equation}
I(\xi, a) = \int_{0}^{\xi} \dfrac{\sin^2\zeta}{\sqrt{\sin^2\zeta+a}}d\zeta,
\end{equation}

The first two parts of the indentation can be easily solved
\begin{equation}
\begin{array}{lcl}
h_{global}&=&R_b - [z(0)-z(\pi/2)]\\[2ex]
{}&=&R_b - \dfrac{\sigma_b}{P}\left[1+I\left(\dfrac{\pi}{2}, a\right)\right];\\[4ex]
h_{dent}&=& z(0)-z(\pi/2-\alpha)\\[2ex]
{}&=& \dfrac{\sigma_b}{P}\left[1-\sin\alpha-I(\pi/2-\alpha, a)\right].\\[2ex]
\end{array}
\end{equation}
$h_{cone}$ is determined by the radius of the contact circle where the normal force between the membrane bulge and the indenter vanishes.  From Equation (\ref{eqn_r}), setting $\theta = \pi/2 - \alpha$, the radius of the contact circle is
\begin{equation}
r_{cone} = \frac{\sigma_b}{P}\left(\sqrt{\cos^2\alpha + a} - \cos\alpha\right),
\end{equation}
so that
\begin{equation}
h_{cone} = \frac{\sigma_b}{P}\left(\sqrt{\cos^2\alpha+a}-\cos\alpha\right)\cot\alpha.
\end{equation}

\section{Radial expansion of an inflated cylinder with stress-stiffening}

Here, we model the radius $R_c$ of an elastic cylinder under variable internal pressure $P$.  Radial expansion of a cylinder under pressure is governed by the elasticity of the cylinder wall in the circumferential direction, $E_\perp$. In the follow discussion, we set the Poisson's ratio to zero.  The circumferential surface tension on the wall
\begin{equation}
\label{eqn_sigma}
\sigma_\perp = PR_c,
\end{equation}
and the natural stress is $\sigma_\perp/t$, where $t$ is the thickness of the wall.  $E_\perp$ is defined using the natural stress and  the incremental strain, $dR_c/R_c$:
\begin{equation}
\label{eqn_def_E}
E_\perp\dfrac{dR_c}{R_c} = \dfrac{d\sigma_\perp}{t}.
\end{equation}

We also assume that $E_\perp$  depends on the internal pressure $P$ and follows a power law
\begin{equation}
\label{eqn_stress_stiffening}
E_\perp=E_0\left(\dfrac{P}{P_0}\right)^\gamma,
\end{equation}
where $E_0$ and $P_0$ can be combined to one single free parameter, $E_0/P_0^\gamma$.  Without loss of generality, we choose $E_0$ as the free parameter and fix $P_0 = 5$~kPa, a typical turgor pressure in a bulging cell.  Let $R_0$ be the radius of the cylinder at pressure $P_0$. We define the following dimensionless quantities
\begin{equation}
\begin{array}{rcl}
\hat{P} &=& P/P_0;\\[2ex]
\hat{E} &=& E_\perp/E_0;\\[2ex]
\hat{R} &=& R_c/R_0;\\[2ex]
\hat{\sigma} &=& \dfrac{\sigma_\perp}{P_0R_0};\\[2ex]
p &=& \dfrac{P_0R_0}{E_0t}.\\
\end{array}
\end{equation}
Equations (\ref{eqn_sigma}), (\ref{eqn_def_E}) and (\ref{eqn_stress_stiffening}) can then be rewritten as
\begin{equation}\label{eqn_sigma_dimless}
\hat{\sigma} = \hat{P}\hat{R},
\end{equation}
\begin{equation}
\hat{E}\dfrac{d\hat{R}}{\hat{R}}=pd\hat{\sigma},
\end{equation}
\begin{equation}
\label{eqn_stress_stiffening_dimless}
\hat{E}=\hat{P}^\gamma,
\end{equation}
Equations  (\ref{eqn_sigma_dimless}-\ref{eqn_stress_stiffening_dimless}) can be combined and solved to yield
\begin{equation}
\dfrac{d\hat{R}}{\hat{R}^{\gamma+1}}=p\dfrac{d\hat{\sigma}}{\hat{\sigma}^\gamma},
\end{equation}
and
\begin{equation}
\label{eqn_sigmaR}
\hat{\sigma} = \left[1+\dfrac{1-\gamma}{p\gamma}\left(1-\hat{R}^{-\gamma}\right)\right]^{\frac{1}{1-\gamma}}
\end{equation}
In the limit of linear stress stiffening, i.e. $\gamma \rightarrow 1$, the dimensionless tension reduces to
\begin{equation}
\hat{\sigma} = \exp\left[\dfrac{1}{p}\left(1-\dfrac{1}{\hat{R}}\right)\right]
\end{equation}
Using Equation (\ref{eqn_sigma_dimless}), we  obtain the desired relationship between the pressure and the inflated radius
\begin{equation}
\frac{P}{P_0} =
 \frac{R_0}{R_c}\left[\frac{(\gamma-1)E_0t}{\gamma P_0R_0}\left[(\frac{R_0}{R_c})^\gamma-1\right]+1\right]^{\frac{1}{1-\gamma}}
\end{equation}
Again, this is simplified in the limit $\gamma \rightarrow 1$:
\begin{equation}
 \frac{P}{P_0} = \frac{R_0}{R_c}\exp\left[\frac{E_0t}{P_0R_0}(1-\frac{R_0}{R_c})\right]
\end{equation}

\section{Scaling laws}
When an inflated cylinder is indented by a conical indenter, the force $F$ required to generate an indentation $h$ can be written as
\begin{equation}
  F=\mathcal{F}(h;E, P, R, t, \delta),
\end{equation}
where $E$ and $P$ are mechanical parameters corresponding to stiffness and pressure and $R$, $t$, and  $\delta$, are  geometric parameters corresponding to the cylinder radius, cylinder thickness and indenter tip radius.  The inflated cylinder radius can be written  as
\begin{equation}
  R=\mathcal{R}(E, P, R_0, t),
\end{equation}
where $R_0$ is the radius at any given reference pressure $P_0$.   The functions $\mathcal{F}$ and $\mathcal{R}$ are determined by the material properties of the material under consideration.  Due to the linearity of solid mechanics, scaling the mechanical parameters results in a scaling of the function $\mathcal{F}$ but leaves $\mathcal{R}$ unchanged:
\begin{equation}
  \mathcal{F}(h;\lambda E, \lambda P, R, t, \delta)= \lambda \mathcal{F}(h;E, P, R, t, \delta);
\end{equation}
\begin{equation}
 \mathcal{R}(\lambda E, \lambda P, R_0, t)=\mathcal{R}(E, P, R_0, t).
\end{equation}
Similarly, when all spatial dimensions scale we have
\begin{equation}
  \mathcal{F}(\mu h; E, P, \mu R, \mu t, \mu \delta)= \mu^2 \mathcal{F}(h;E, P, R, t, \delta);
\end{equation}
\begin{equation}
\mathcal{R}(E, P, \mu R_0, \mu t)=\mu \mathcal{R}(E, P, R_0, t).
\end{equation}
For thin shells, one additional scaling rule applies according to Kirchhoff--Love theory \cite{Love1888}:
\begin{equation}
  \mathcal{F}(h; \eta E, P, R, \eta^{-1}t, \delta)= \mathcal{F}(h;E, P, R, t, \delta);
\end{equation}
\begin{equation}
\mathcal{R}(\eta E, P, R_0, \eta^{-1}t)=\mathcal{R}(E, P, R_0, t).
\end{equation}
The three scaling laws together reduce the total number of independent parameters in the cylinder--indentation problem:
\begin{equation}
\label{eqn_Fscale}
  \mathcal{F}(h; E, P, R, t, \delta)= PR^2\cdot \mathcal{F}(h/R; Et/PR, 1, 1, 1, \delta/R);
\end{equation}
\begin{equation}
\label{eqn_Rscale}
1=\mathcal{R}(Et/PR, 1, R_0/R, 1).
\end{equation}
From Equation (\ref{eqn_Fscale}), we obtain the scaling rule for the stiffness:
\begin{equation}
  \dfrac{d\mathcal{F}(h; E, P, R, t, \delta)}{dh}= PR\cdot \mathcal{F}'(h/R; Et/PR, 1, 1, 1, \delta/R).
\end{equation}
For small indentations, where the stiffness of the material can be considered to be nearly linear, $\mathcal{F}'$ is independent of the indentation depth $h$.  Therefore, $k/PR$ only depends on $PR/Et$ and the scaled indenter size $\delta/R$.  In addition, from Equation (\ref{eqn_Rscale}), the dependence of $R/R_0$ on $PR/Et$ can be implicitly solved.

\section{Anisotropy of the elastic modulus in the presence of stress-stiffening}

For a cylinder, the surface tension is anisotropic. The tension in the  circumferential direction is  twice that along the axial direction. For a stress-stiffening material, this results in an anisotropic elasticity. Here, we find the ratio of the axial elasticity to the circumferential elasticity at a given pressure.

Equations (\ref{eqn_sigma_dimless}) and (\ref{eqn_sigmaR}) can be combined to give the dimensionless pressure at a given surface tension
\begin{equation}
\hat{P}(\hat{\sigma}) = \hat{\sigma}\left[1-\dfrac{p\gamma}{1-\gamma}\left(\hat{\sigma}^{1-\gamma}-1\right)\right]^{\frac{1}{\gamma}}
\end{equation}
which reduces to
\begin{equation}
	\hat{P}(\hat{\sigma})  = \hat{\sigma}\left(1-p\ln\hat{\sigma} \right)
\end{equation}
in the limit that $\gamma \rightarrow 1$. The elastic modulus is found by taking this expressions to the power $\gamma$:
\begin{equation}
\label{eqn_Eonsigma}
\hat{E}(\hat{\sigma}) = \hat{\sigma}^\gamma\left[1-\dfrac{p\gamma}{1-\gamma}\left(\hat{\sigma}^{1-\gamma}-1\right)\right]
\end{equation}
Note that Equation (\ref{eqn_Eonsigma}) is a general expression that applies in both the circumferential and axial directions, where the surface tension differs by a factor of two.  Therefore, the anisotropic ratio of the two elasticities can be found as
\begin{equation}
  \dfrac{E_\parallel (\hat{P})}{E_\perp(\hat{P})} = \dfrac{\hat{E}\left[\hat{\sigma}_c(\hat{P})/2\right]}{\hat{E}\left[\hat{\sigma}_c(\hat{P})\right]}.
\end{equation}

\newpage
\section{List of symbols}
\begin{table}[h]
\caption{List of constants and symbols}
\centering
\begin{tabular}{|c|l|l|}
\hline
symbol & description & value\\
\hline
 \hline
  $\alpha$ & indenter half--cone angle & $\pi/12$\\
  \hline
  $a$ & normalized indentation force & $PF / \pi \sigma_b^2$\\ 
  \hline
  $P_0$ & normalization constant for pressure & 5000 Pa\\
  \hline
  $t$ & cell wall thickness & $4.5\pm1.5$~nm\\
  
\hline
  $E_0$ & circumferential Young's modulus of the cell wall at pressure $P_0$&\\
  \hline
  $E_\perp$ & circumferential Young's modulus of the cell wall& \\
    \hline
  $E_\parallel$ & axial Young's modulus of the cell wall& \\
  \hline
    $F$ & indentation force&\\
          \hline
  $\gamma$ & stress--stiffening exponent&\\
      \hline
  $h$ & total deformation&\\
  \hline
    $k_b$ & bulge indentation stiffness&\\
    \hline
  $k_c$ & cell indentation stiffness&\\
      \hline
  $P$ & turgor pressure&\\
  \hline
  $R_b$ & bulge radius&\\
    \hline
  $R_c$ & cell radius&\\
      \hline
  $\sigma_b$ & surface tension of the bulge&\\
    \hline
  $\sigma_\perp$ & cell wall circumferential surface tension&\\
  \hline
  $\sigma_\parallel$ & cell wall axial surface tension&\\
    \hline
  $\theta$ & elevation angle along bulge&\\
    \hline
  $(r, z)$ & cylindrical  coordinates in bulge calculation&\\
  \hline
  $R_0$ & cell radius at pressure $P_0$&\\
  \hline
  IM & inner membrane&\\
  \hline
  OM & outer membrane&\\
  \hline
  PG & peptidoglycan&\\
  \hline
\end{tabular}
\end{table}

\newpage
%\bibliographystyle{is-unsrt}
%\bibliography{YD_PRL_supp}